\documentclass[pre,twocolumn,aps,amsmath,amssymb]{revtex4}
\usepackage{amsfonts}
\usepackage{bbold}
\usepackage{mathrsfs}
\usepackage{color}
\usepackage{epstopdf}

\usepackage{graphicx}% Include figure files
\usepackage{bm}%bold greek letters
\usepackage{amssymb, dsfont}

\newcommand{\nn}{\nonumber }

\newcommand{\BEQ}{\begin{equation}}
\newcommand{\EEQ}{\end{equation}}
\newcommand{\BEA}{\begin{eqnarray}}
\newcommand{\EEA}{\end{eqnarray}}

\begin{document}

%\preprint{}

\title{The effect of time-correlated noise on the Kuramoto model\\ studied via the unified colored noise approximation}% Force line breaks with \\

\author{Claudio Maggi$^1$}
\affiliation{$^1$NANOTEC-CNR, Institute of Nanotechnology, Soft and Living Matter Laboratory, Rome, Italy}
\email{claudio.maggi@roma1.infn.it }

\author{Matteo Paoluzzi$^2$}
%\affiliation{$^2$Dipartimento di Fisica, Universit\'{a} di Roma ``Sapienza'', I-00185, Rome, Italy}
\affiliation{Institute of Complex Systems (ISC-CNR) and Dipartimento di Fisica, Sapienza University of Rome, I-00185 Rome, Italy.}
\email{Matteo.Paoluzzi@roma1.infn.it}

\begin{abstract}
Many natural and social phenomena are characterized by synchronization. 
The Kuramoto model, taking into account the basic ingredients for observing synchronized states,
allows to study mathematically synchronization  in a simplified
but nontrivial picture. Here we study how a noise that is correlated on a finite 
time-scale $\tau$ impacts the ability of the Kuramoto model to achieve synchronization.
We develop an approximated theory that allows to compute the critical coupling constant $k_c$ as a function of the correlation time $\tau$. We obtain that that $k_c(\tau)$ decreases as $\tau$ increases indicating that time-correlated noise promotes synchronization. Moreover, we show that theory describes qualitatively well the degree of synchronization near $k_c$ obtained numerically. Finally, we show that, independently on the value of $\tau$, the curves of the order parameter versus $k$ scale on the same master curve even at values of $k$ very far from $k_c$.
\end{abstract}

\maketitle

\section{Introduction}

The firing of neurons in the visual cortex \cite{eckhorn1988coherent,gray1989oscillatory},
the frequency locking in Josephson arrays \cite{PhysRevE.57.1563}, and the flashing 
in large groups of fireflies \cite{buck1968mechanism}
are just a few examples, ranging
from biology to physics, where synchronization plays a fundamental role \cite{strogatz2003sync}. 
Winfree realized that nonlinear interactions are a key ingredient for synchronization phenomena
that happen above a threshold value of the coupling constant \cite{winfree1967biological,winfree2001geometry}.
The Kuramoto model (KM), originally introduced by Kuramoto in 1975, contains all the basic ingredients needed to observe synchronization \cite{kuramoto1975self}. 
  In the KM a population of oscillators, each of them characterized by its natural frequency, are
 coupled globally through a non-linear interaction. The model is thus general enough to describe diverse situations,
 contains a small number of control parameters, and, because of the mean-field coupling, is analytically tractable.

 The KM undergoes a continuous phase transition towards a partially synchronous state \cite{acebron2005kuramoto} as the control parameter of the transition (the coupling constant $k$ between the oscillators) is increased above a critical value. When white noise is taken into account, this acts as a random perturbation that prevents the system from reaching a perfectly synchronized state and makes the critical value of the coupling constant noise-dependent, i. e., $k_c=k_c(T)$, with the ``temperature'' $T$ representing the  strength of the noise. However, in many biological systems \cite{PhysRevLett.82.2402}, the external noise is not characterized by a flat spectrum. 
 A noise source that is exponentially correlated on a finite time scale $\tau$, is a step towards a more realistic description of the system. 
In this paper, we study the KM in the presence of a Gaussian 
exponentially correlated noise \cite{tonjes2010synchronization,bag2007influence,Moro_1998}. Following the standard terminology in the field of noise-driven dynamical systems~\cite{hanggi1995p}, we refer to this type of noise as \textit{Gaussian colored noise} (the adjective ``colored'' refers to the fact the power spectrum of the noise is not flat/white as the noise spectrum of standard Brownian motion).

We show that, within the Unified Colored Noise Approximation
(UCNA) \cite{praucna,hanggi1995p,mucna}, it is possible to 
solve the colored-noise driven KM for the critical coupling constant $k_c$ which becomes a decreasing function of $\tau$. This is in agreement with early numerical and theoretical works \cite{bag2007influence,Moro_1998} and we further test it by computer simulations which, however, reveal significant quantitative deviations from the UCNA predictions at large values of $\tau$. Moreover, we compute analytically 
the critical behavior of the order parameter that is found to grow as $[ k - k_c(\tau) ]^{1/2}$, i.e. with the same mean-field exponent of the KM in presence of thermal noise. Finally, we show that, intriguingly, the synchronization order parameter of the KM displays a universal behavior that holds well even when $k$ is very far from $k_c$, at all values of $\tau$ simulated.

\section{Theory}

We consider the KM defined by the set of equations~\cite{acebron2005kuramoto}

\begin{equation} \label{eqm}
\dot{\phi}_i = \omega_i - \frac{k}{N} \sum_{j=1}^N \sin(\phi_i-\phi_j) + \eta_i
\end{equation}

\noindent where $\phi_i \in [0,2\pi]$ is the phase of the $i-$th oscillator ($i=1,...,N$), $k$ is the coupling strength and the $\omega_i$ is the (random) natural frequency distributed among the oscillators according to the probability function $g(\omega)$ which is assumed to be symmetric and with zero mean. The $\eta_i$ are a set colored noise sources evolving according to the stochastic equations:

\begin{equation} \label{neqm}
\dot{\eta}_i = -\tau^{-1} \eta_i + D^{1/2} \tau^{-1} \zeta_i
\end{equation}

\noindent where $\tau$ is the correlation time of the noise, $D$ is the noise amplitude and the $\zeta_i$ are a set of uncorrelated standard white noise sources: ${\langle \zeta_i(0) \zeta_j(t)\rangle = 2 \delta_{ij} \, \delta(t) }$. 

Following Ref.~\cite{sakaguchi1988cooperative} we define the (complex) order parameter $\sigma \, \exp(i\,\phi_0)$ as

\begin{equation} \label{op}
\sigma \, \exp(i\,\phi_0) = 
\frac{1}{N}\sum_{j=1}^N \exp(i\,\phi_j)
\end{equation}

\noindent where the amplitude $\sigma$ of the order parameter can be interpreted as the degree of synchronization of the oscillators population ($0 \leq \sigma \leq 1$). By using Eq.~(\ref{op}) we can rewrite Eq.~(\ref{eqm}) as

\begin{equation} \label{eqm2}
\dot{\psi} = \omega - k \, \sigma \, \sin \psi + \eta
\end{equation}

\noindent where $\psi=\phi_i-\phi_0$. 
Since from this point we focus on one single oscillator (the $i$-th oscillator), we have dropped the index $i$ for simplicity (setting ${\eta_i=\eta}$ and ${\omega_i=\omega}$) as in Ref.~\cite{sakaguchi1988cooperative}.

It is well known that~\cite{hanggi1995p}, for non-linear colored noise-driven systems as the one described by Eq.~(\ref{eqm2}), further progress can be made only by using some approximation scheme.
We now follow ref.~\cite{hanggi1995p} to illustrate how the UCNA applies to our problem. We first write Eqs.~(\ref{eqm2}) and~(\ref{neqm}) as one single equation of higher order

\begin{equation} \label{soeq}
\ddot{\psi} + \dot{\psi}\, [\tau^{-1}-f'(\psi)]-f(\psi)/\tau= D^{1/2} \zeta/\tau
\end{equation}

\noindent
where $f(\psi)=\omega - k \, \sigma \, \sin \psi$ is the ``mean-field force'' appearing in Eq.~(\ref{eqm2}) and the prime indicates differentiation with respect to $\psi$. Eq.~(\ref{seq}) shows that the dynamics of the phase $\psi$ is governed by the field $f$, the white noise $\zeta$ and the non-homogeneous friction $[\tau^{-1}-f(\psi)]$. It is now convenient to introduce the time-scale $s=\tau^{-1/2}t$, so that Eq.~(\ref{soeq}) becomes

\begin{equation} \label{seq2}
\ddot{\psi} + \dot{\psi}\, [\tau^{-1/2}-\tau^{1/2}
f'(\psi)]-f(\psi) = D^{1/2} \zeta/\tau^{1/4} \; .
\end{equation}

\noindent
The UCNA consists in eliminating adiabatically the term $\ddot{\psi}$ in Eq.~(\ref{seq2}) which is justified whenever $\tau$ is small. If $\tau$ is large the validity of the approximation is restricted to the case where $f'(\psi)$ is negative and large, i.e. when the external potential $u(\psi)=-\int^\psi d\psi f(\psi)$ has high positive curvature. 
In the present case 
$u(\psi)$ 
has curvature $u''(\psi)=\sigma \, k\, \cos(\psi)$, which is negative for $\pi/2<\psi<3\pi/2$, so we expect that the UCNA is accurate only for small values of $\tau$ (as confirmed by numerical simulations in the next section).
Having neglected the inertial term $\ddot{\psi}$ in Eq.~(\ref{seq2}), and going back from the time variable $s$ to $t$, we get the final UCNA stochastic evolution equation

\begin{equation} \label{seq3}
\dot{\psi} = [1-\tau \, f'(\psi)]^{-1}f(\psi) +
[1-\tau \, f'(\psi)]^{-1}D^{1/2} \zeta \; .
\end{equation}

%\\
\noindent
Crucially, from Eq.~(\ref{seq3}) we can write the Fokker-Planck equation~\cite{risken1996fokker} determining the evolution of the phase probability distribution
$n = n(t,\psi,\omega)$

\begin{equation} \label{eqmp}
\dot{n} = -\partial_\psi \left[
f \, \gamma \, n  - D \, \gamma \, 
\partial_\psi (\gamma \, n)
\right]
\end{equation}

\noindent 
where $\gamma = [1-\tau \partial_\psi f ]^{-1}$. The stationary probability distribution $n(\psi,\omega)$ is found from Eq.~(\ref{eqmp}) by setting the probability current to a constant:

\begin{equation} \label{eqst}
f \, \gamma \, n  - D \, \gamma \, 
\partial_\psi (\gamma \, n) = \mathrm{const} \, .
\end{equation}

\noindent The solution of Eq.~(\ref{eqst}) can be obtained by imposing periodic the boundary condition $n(0,\omega)=n(2 \pi,\omega)$. Following the method of  Ref.~\cite{marconi2017self} we thus have

\begin{widetext}

\begin{eqnarray}
n(\psi,\omega) &=& \mathcal{N} \;
\frac{e^{F(\psi,\omega)} \, G(\psi)}{1-e^{2 \pi \, \omega/D}}
\left( 
\int_\psi^0 d\psi' e^{-F(\psi',\omega)}\, G(\psi')
+ e^{2 \pi \, \omega/D}
\int_{2\pi}^\psi d\psi' e^{-F(\psi',\omega)} \, G(\psi')
\right) \\ \nn \textrm{where}&& \\
\nn
F (\psi,\omega) &\equiv& \frac{k \, \sigma \,  [k \, \sigma \,  \tau \,  \cos (2 \psi )+4 \, \tau \,  \omega  \sin \psi
   +4 \cos \psi ]+4 \psi  \, \omega }{4 D} \\ \nn
G (\psi) &\equiv& |1+k \, \sigma \, \tau \,  \cos \psi|
\end{eqnarray}

\end{widetext}
%
%\noindent where
%
%\begin{eqnarray*}
%&& F (\psi,\omega) = \frac{k \, \sigma \,  [k \, \sigma \,  \tau \,  \cos (2 \psi )+4 \, \tau \,  \omega  \sin \psi
%   +4 \cos \psi ]+4 \psi  \, \omega }{4 D} \\
%&& G (\psi) = |1+k \, \sigma \, \tau \,  \cos \psi|
%\end{eqnarray*}

\noindent and the constant $\mathcal{N}$ is set by the normalization condition $\int d \psi \, n(\psi,\omega)=1$. The self-consistent equation for the order parameter $\sigma$ is obtained by integrating over the ``disorder'' $\omega$, i.e. by integrating over all oscillators each characterized by a natural frequency $\omega$

\begin{equation} \label{selfc}
\sigma = \int_{-\infty}^{\infty} d \omega \, g(\omega) 
\int_0^{2\pi} d \psi \, n(\psi,\omega) \, \exp(i \, \psi) \; .
\end{equation}

\noindent To find the critical coupling strength $k_c$ and the behavior of $\sigma$ near $k_c$ we expand the r.h.s. of Eq.~(\ref{selfc}) up to $\sigma^3$ so that this equation becomes

\begin{equation} \label{selfce}
\sigma 
- k \, \sigma \int_{-\infty}^\infty d\omega  \, Q(\omega)
+ k^3 \, \sigma^3 \int_{-\infty}^\infty d\omega  \, P(\omega)
=0
\end{equation}

\noindent where

\begin{widetext}

\begin{eqnarray*}
&& Q(\omega) = \frac{1}{2} g(\omega ) \left(\frac{D}{D^2+\omega ^2}+\tau \right)\\
&& P (\omega) = \frac{g(\omega)
\left[ 2 D^3-4 D \omega ^2+\tau ^2 \left(6 D^5+6 D^3 \omega ^2\right)+\tau  \left(4 D^4+5 D^2 \omega ^2+\omega
   ^4\right) \right]}{8 \left(D^2+\omega ^2\right)^2 \left(4 D^2+\omega ^2\right)} \; .
\end{eqnarray*}

\end{widetext}

\noindent Obviously, $\sigma=0$ is always a solution of Eq.~(\ref{selfce}). However, an additional real positive root of the form

\begin{equation} \label{seq}
\sigma = \left[  \frac{k \,\int d\omega \, Q(\omega) \; - 1}{k^3\int d \omega P(\omega) }\right]^{1/2}
\end{equation}

\noindent appears when ${k \int d\omega  \, Q(\omega)>1}$ yielding the equation ${k_c=1/\int d\omega \, Q(\omega)}$, i.e.

\begin{equation} \label{kc}
k_c(\tau) =\frac{2}{\tau + D \int_{-\infty }^{\infty } d\omega
\frac{ g(\omega )}{D^2+\omega ^2} }
\end{equation}

\noindent
Note that Eq.~(\ref{kc}) suggest that $k_c$ is a decreasing function of $\tau$ and reduces to the well known formula for the critical coupling of the KM in presence of white noise in the limit $\tau \rightarrow 0$ (see Ref.s \cite{sakaguchi1988cooperative}
and \cite{bag2007influence}).

\begin{figure}[ht]
  \centering
  \includegraphics[width=.95\columnwidth]{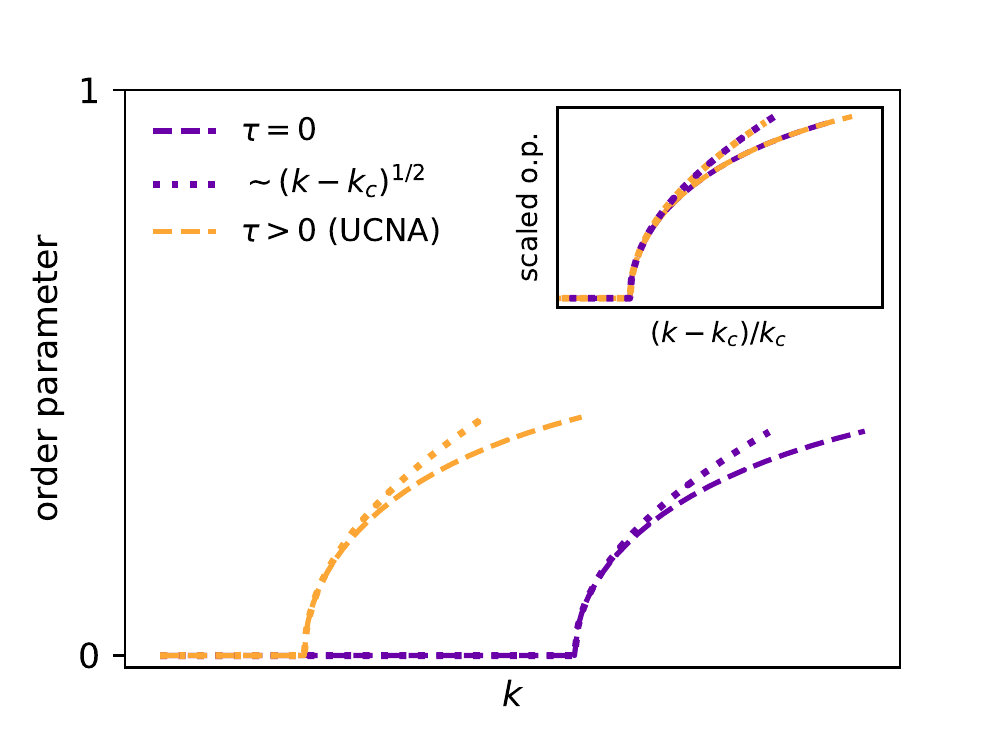}
  \caption{Sketch of the UCNA scenario for the degree of synchronization $\sigma$ of the colored noise-driven KM. The main panel shows how the $\sigma(k)$ curve (Eq.~(\ref{seq})) shifts to lower $k$ values for $\tau>0$ as $k_c$ decreases (dashed curves, see legend). The dotted curves indicates the approximated $\sigma(k)$ in the critical regime (Eq.~(\ref{scrit})). The inset shows the same curves of the main panel as a function of the reduced coupling constant and scaled according to Eq.~(\ref{seqsc}), it is evident that the scaling holds beyond the critical regime.}
  \label{f0}
\end{figure}

It is also interesting to note that 
the equation for the order parameter derived within the UCNA (Eq.~(\ref{seq})) depends on $\tau$ only via the integrals of $P$ and $Q$ while keeping the $k$-dependence untouched.
In practice $\sigma(k)$ at $\tau>0$ is a shifted and scaled version of the standard $\sigma(k)$ curve of the white-noise driven KM as sketched in Fig.~\ref{f0}.

This implies that $\sigma$ is a ``universal'' function of $k$ when properly scaled. To see this let us introduce the reduced coupling constant $\kappa = [k-k_c(\tau)]/k_c(\tau)$ so that Eq.~(\ref{seq}) becomes

\begin{equation} \label{seqsc}
\sigma = \mathcal{A(\tau)} 
\left[ \frac{\kappa}{(1+\kappa)^3} \right]^{1/2}
\end{equation}

\noindent
where ${\mathcal{A}(\tau)=(\int d \omega \, Q )^{3/2}/(\int d \omega \, P)^{1/2}}$. It is evident from Eq.~(\ref{seqsc}) that the scaled order parameter $\sigma/\mathcal{A}$ as a function of $\kappa$ is independent on $\tau$. We stress that the universal functional form of Eq.~(\ref{seqsc}) is expected to be valid for small $\kappa$, however the validity of Eq.~(\ref{seqsc}) is not restricted only to ``critical'' regimes where $k$ is extremely close to $k_c$ but it should hold also for larger values of $\kappa$ (see the inset of Fig.~\ref{f0}). In the near-critical regime, $\sigma$ can be further approximated by

\begin{equation} \label{scrit}
\sigma \approx 
\mathcal{A}(\tau) \, \kappa^{1/2}
\end{equation}

Summarizing our results, by using the UCNA, we have found that, while the critical coupling strength decreases with $\tau$ (Eq.~(\ref{kc})), the behavior of the order parameter near $k_c$ should be qualitatively unaffected by the memory of the noise 
(Eq.~(\ref{seqsc})) and also the mean-field critical exponent $1/2$ should remain the same as the one of the white-noise driven KM (Eq.~(\ref{scrit})). 

\section{Simulations}

In order to test the results of the previous section, we integrate numerically the equations of motion (\ref{eqm}) and (\ref{neqm}) by using an Euler algorithm that, taking advantage of the peculiar form of the interaction, results in a computational cost $\mathcal{O}(N)$~\cite{januszewski2010accelerating}. Following previous works~\cite{bag2007influence,tonjes2010synchronization} we simulate $N=5000$ oscillators fixing $D=1$ scanning several values of $k$ and $\tau$. We make the usual choice of distributing the $\omega_i$ according to a Cauchy-Lorentz probability function 

\begin{equation} \label{gw}
g(\omega) = \frac{1}{\pi}\frac{\lambda}{\omega^2 + \lambda^2} 
\end{equation}

\noindent setting $\lambda=1/2$. In all simulations the time-step is set to $10^{-3}$, each simulation runs at fixed $k$,$\tau$ and $D$ for $2\times 10^6$ steps and the order parameter $\sigma$ is computed as in Eq.~(\ref{op}) over the last $10^6$ steps averaging over all spins.

\begin{figure}[ht]
  \centering
  \includegraphics[width=.9\columnwidth]{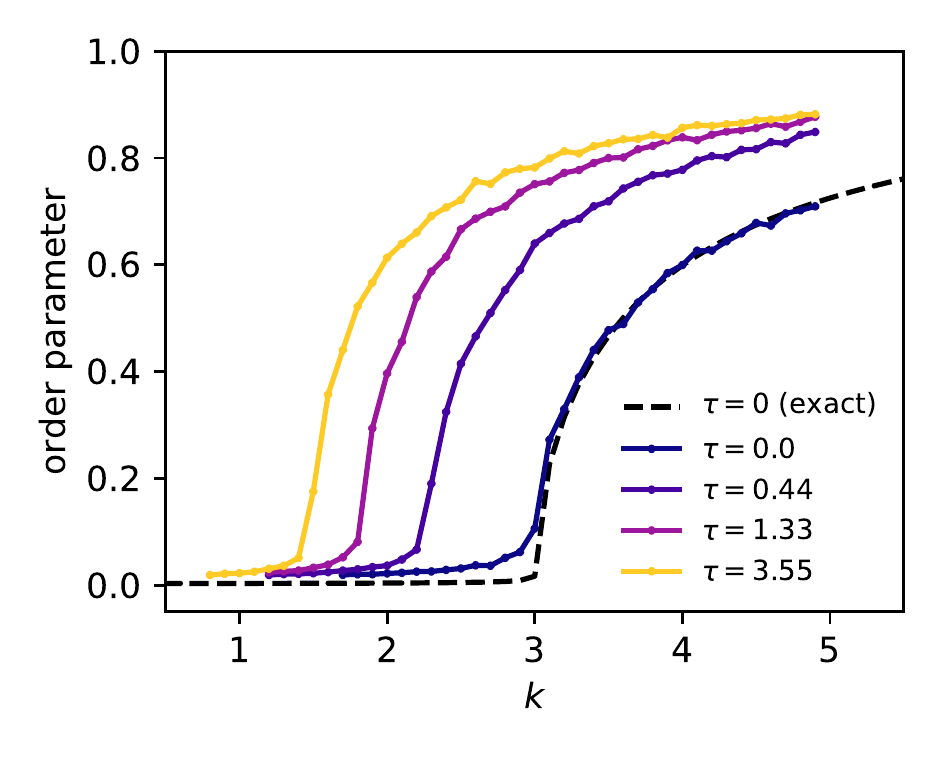}
  \caption{Numerical values of the order parameter for a system of $N=5000$ oscillators as a function of $k$. Each colored curve corresponds to a different value of $\tau$ (see legend). The dashed line is the numerical solution of the self-consistent equation for $\tau=0$.}
  \label{f1}
\end{figure}

Simulation results are shown in Fig.~\ref{f1} where it is clear that the order parameter goes to zero at $k$ values that become progressively lower as $\tau$ increases. 
In Fig.~\ref{f1} we also compared the simulation data with the numerical solution of the theoretical Eq.~(\ref{selfc}) for $\tau=0$. To obtain this solution we evaluate the integrals in Eq.~(\ref{selfc}) by the trapezoidal rule and solve at each value of $k$ by using the Newton method.
The close agreement between the simulations and  Eq.~(\ref{selfc}) confirms that a system with $N=5000$ behaves essentially as in the thermodynamic limit (as noticed \textit{e.g.} in Ref.~\cite{bag2007influence}) with some deviation observed for $k<k_c$.

\begin{figure}[ht]
  \centering
  \includegraphics[width=.9\columnwidth]{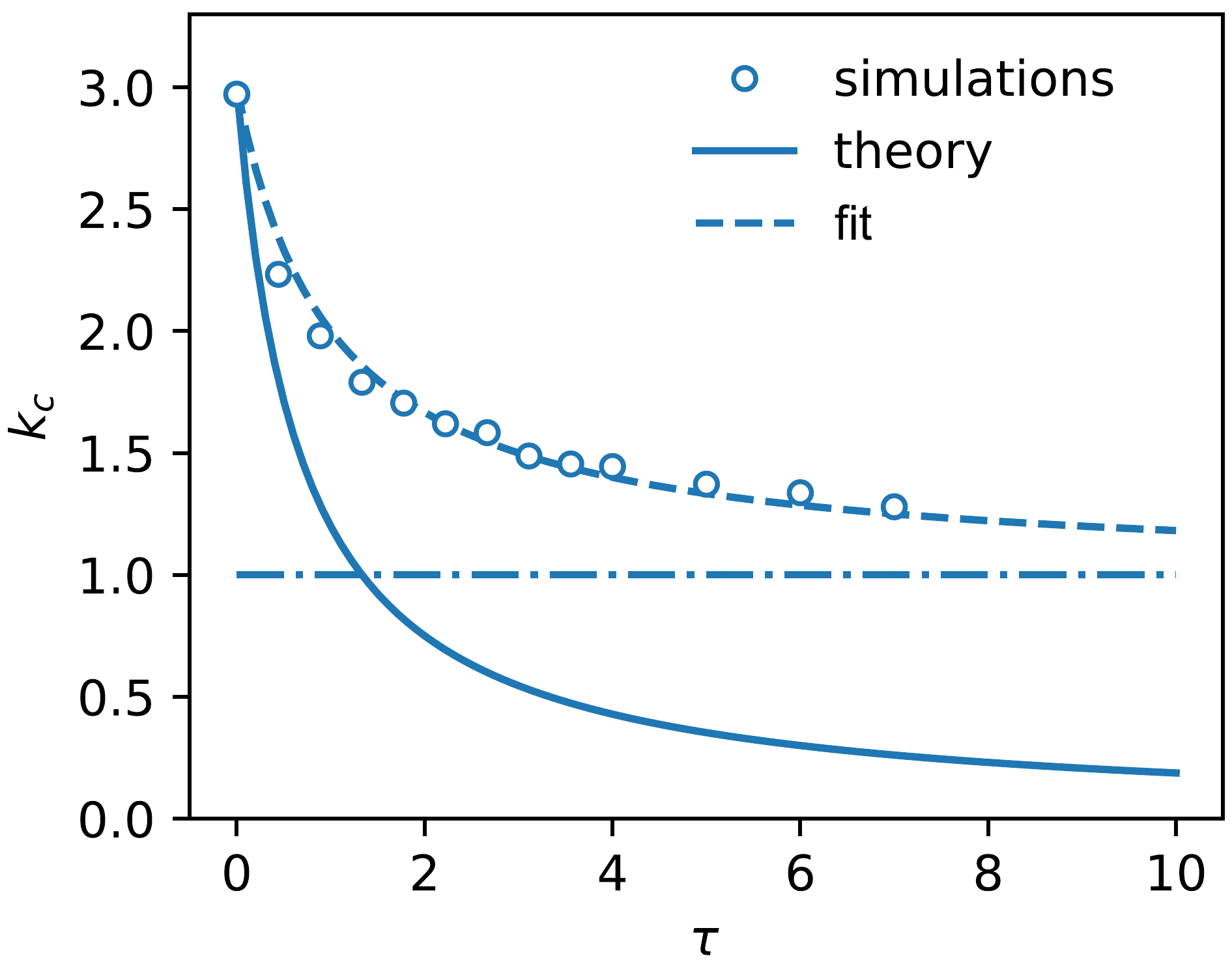}
  \caption{Symbols are the values of $k_c$ obtained numerically. The full line is the theoretical prediction of the UCNA. The dashed dotted line is the expected asymptote of $k_c(\tau)$ in the limit $\tau \rightarrow \infty$. The dashed line is a fit with an empirical formula.}
  \label{f2}
 \end{figure}

From Fig.~\ref{f1} follows that $k_c$ is a decreasing function of $\tau$ as predicted by the theory (see Eq.~(\ref{kc})). However, a more accurate comparison shows clear quantitative deviations. To see this, we calculate $k_c(\tau)$ from Eq.~(\ref{kc}) with the choice (\ref{gw}) 

\begin{equation} \label{kcl}
k_c(\tau) = \frac{2 (D+\lambda )}{1+\tau  (D+\lambda )}
\end{equation}

\noindent and compare this formula with the values of $k_c$ obtained numerically in Fig.~\ref{f2}. We see that Eq.~(\ref{kcl}) is close to the numerical $k_c$ for small $\tau$ yielding the exact result 
$k_c(0)=2 (D+\lambda)$ at $\tau=0$, however it quickly deviates from the numerical data as $\tau$ increases. 

In particular, while Eq.~(\ref{kcl}) predicts that $\lim_{\tau \rightarrow \infty} k_c(\tau) = 0$, the numerical value of $k_c$ approaches an asymptotic value $k_c(\infty)>0$. This behavior is expected if we consider that the colored noise ``intensity'' $\langle \eta^2 \rangle = D/\tau$ (from Eq.~(\ref{neqm})) goes to zero in the $\tau \rightarrow \infty$ limit at fixed $D$.
Therefore we expect that, in this limit, the critical coupling coincides with the $k_c=2\lambda$ of the KM in absence of noise~\cite{bag2007influence}. Moreover, the breakdown of the UCNA for large $\tau$ values is expected from the discussion of the previous section: when the inertial term in the Langevin equation (\ref{seq2}) cannot be neglected the mapping of the original colored noise problem into a white noise-driven dynamical system fails at large $\tau$.

To underline even more how the UCNA quantitatively deviates from the simulations we fit the $k_c$ data with a simple empirical formula (see Fig.~\ref{f2}).  Assuming that the actual $k_c(\tau)$ has the form of a hyperbola and requiring that this coincides with the known limiting values

\begin{eqnarray}
&&k_c(\tau \rightarrow 0) = 2(D+\lambda)\\
&&k_c(\tau \rightarrow \infty) = 2\lambda \nonumber 
\end{eqnarray}

\noindent
we fit the simulation data points with

\begin{equation}
k_c(\tau) = k_c(\infty) + \frac{k_c(0)-k_c(\infty)}{1+\alpha \, \tau}
\end{equation}

\noindent
where $\alpha$ is an adjustable parameter. In Fig.~\ref{f2} we show that such a simple equation interpolates well the data (with $\alpha\approx0.5$) signaling that, while the UCNA leads to a quantitatively wrong result, the functional form of $k_c(\tau)$ is still well captured by a hyperbola.

  \begin{figure}[ht]
  \centering
  \includegraphics[width=.95\columnwidth]{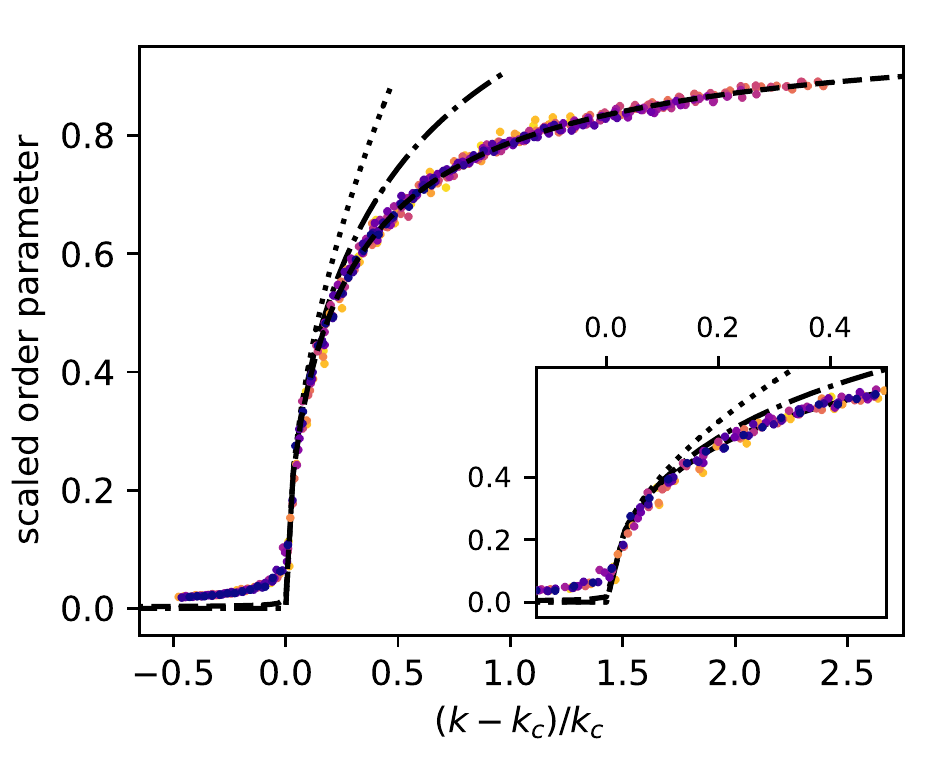}
  \caption{Colored dots represents the scaled order parameter as function of the scaled coupling strength. Each color represent simulation data at different values of $\tau$ between $\tau=0$ (blue) and $\tau=7$ (yellow) using the same coloring of Fig.~\ref{f1}. The dashed line is the numerical solution of the self-consistent equation for $\tau=0$. The dashed-dotted line is the approximated analytical solution (Eq.~(\ref{seqsc})). The dotted line is the critical power-law of Eq.~(\ref{scrit}). The inset shows a zoom of the main panel data in the near-critical region.}
  \label{f3}
\end{figure}

Finally, we discuss the scaling of the order parameter $\sigma$ near the transition. As shown in Eq.~(\ref{seqsc}) when $k-k_c(\tau)$ is small the critical behavior of $\sigma$ should be invariant with respect to the correlation time of the noise $\tau$. Therefore all data-points, collected varying $\tau$, should collapse in the same power-law when properly scaled. To test this in Fig.~\ref{f3} we plot the scaled $\sigma$ as a function of the scaled coupling $[k-k_c(\tau)]/k_c(\tau)$ (using the $k_c$ obtained numerically). To scale the $y$-value in this plot we minimize numerically the difference between the simulation values of $\sigma$ and the theoretical curved $\sigma(k-k_c)$ obtained by solving numerically Eq.~(\ref{selfc}) for $\tau=0$ (also shown in Fig.~\ref{f1}) in the near-critical region $(k-k_c)/k_c < 0.5$. 
Quite surprisingly we find that a very good data collapse is observed even far from the critical region (i.e. up to values of $(k-k_c) \approx 2 k_c$). 
To visualize this we also plot Eq.~(\ref{seqsc}) in Fig.~\ref{f3} and we see that a good data collapse is found even where Eq.~(\ref{seqsc}) deviates substantially from the data points.
Justifying this observation from a theoretical point of view is, however, very challenging (even within the simplification of the UCNA) since one in principle should solve analytically the transcendental self-consistency equation defining the order parameter (Eq.~(\ref{selfc}))

In conclusions simulations show that the prediction of the UCNA is qualitatively correct but quantitatively inaccurate for locating the critical coupling strength in presence of colored noise. The scaling of the order parameter predicted by the theory is confirmed numerically and it is found to extend well beyond the expected regime of validity suggesting that the functional form of the order parameter  $\sigma(k)$ is nearly invariant with respect to the correlation time of the noise for all values of $\tau$.

\section{Conclusions}
In this paper, we have studied the KM driven by exponentially correlated noise. We have shown that the UCNA allows to obtain analytical predictions that are in qualitatively agreement
with numerical simulations. In particular, we computed the critical coupling $k_c(\tau)$ that results to be a decreasing function of the correlation time of the noise $\tau$. This finding is confirmed by numerical data that clearly show the increasing of magnitude of the order parameter $\sigma$ as $\tau$ increases. The observed phenomenology is in agreement with early works on similar models \cite{bag2007influence,tonjes2010synchronization,Moro_1998}. Moreover, the theory predicts that colored noise does not change the critical behavior of the order parameter $\sigma$, as confirmed by numerical simulations that
provide evidence for an even more general universal scaling of $\sigma$ as a function scaled coupling $(k-k_c)/k_c$.

What we learn in general from this work is that a persistent noise helps the system to reach a more synchronized state (without affecting the universality class of the model). More specifically we find that the UCNA allows to derive analytically this general result. A similar scenario has been also observed in a closely related lattice-model driven out-of-equilibrium by correlated noise, i.e. the $xy$-model in the presence of a Gaussian colored noise \cite{PRExy}. The picture is also similar to the one of a colored-noise driven $\phi^4$ theory where, however, it was found that the critical coupling does not change monotonically with $\tau$~\cite{PREphi4}. Finally, we want to stress that the fact that these models remain qualitatively identical to the corresponding white-noise driven systems is not a general feature of all colored-noise driven systems. 
This picture differs substantially, for example, from the one found in active repulsive particle systems where the color of noise completely changes the nature of the effective interactions~\cite{farage2015effective,marconi2016effective} and can induce phase transitions that are not present at $\tau=0$~\cite{cates2015motility}. In this context, the UCNA appears as an invaluable tool to understand qualitatively the impact of the noise memory on the phase behavior.

%\textcolor{blue}{
%We hope this work will stimulate future theoretical and experimental investigation
%
%}

\section*{Acknowledgments}
MP acknowledges funding from Regione Lazio, Grant Prot. n. 
85-2017-15257 ("Progetti di Gruppi di Ricerca - Legge 13/2008 - art. 4").
This work was also supported by the Joint Laboratory on ``Advanced and Innovative Materials'', ADINMAT, WIS-Sapienza (MP).

\bibliography{references}

\begin{thebibliography}{25}
\expandafter\ifx\csname natexlab\endcsname\relax\def\natexlab#1{#1}\fi
\expandafter\ifx\csname bibnamefont\endcsname\relax
  \def\bibnamefont#1{#1}\fi
\expandafter\ifx\csname bibfnamefont\endcsname\relax
  \def\bibfnamefont#1{#1}\fi
\expandafter\ifx\csname citenamefont\endcsname\relax
  \def\citenamefont#1{#1}\fi
\expandafter\ifx\csname url\endcsname\relax
  \def\url#1{\texttt{#1}}\fi
\expandafter\ifx\csname urlprefix\endcsname\relax\def\urlprefix{URL }\fi
\providecommand{\bibinfo}[2]{#2}
\providecommand{\eprint}[2][]{\url{#2}}

\bibitem[{\citenamefont{Eckhorn et~al.}(1988)\citenamefont{Eckhorn, Bauer,
  Jordan, Brosch, Kruse, Munk, and Reitboeck}}]{eckhorn1988coherent}
\bibinfo{author}{\bibfnamefont{R.}~\bibnamefont{Eckhorn}},
  \bibinfo{author}{\bibfnamefont{R.}~\bibnamefont{Bauer}},
  \bibinfo{author}{\bibfnamefont{W.}~\bibnamefont{Jordan}},
  \bibinfo{author}{\bibfnamefont{M.}~\bibnamefont{Brosch}},
  \bibinfo{author}{\bibfnamefont{W.}~\bibnamefont{Kruse}},
  \bibinfo{author}{\bibfnamefont{M.}~\bibnamefont{Munk}}, \bibnamefont{and}
  \bibinfo{author}{\bibfnamefont{H.}~\bibnamefont{Reitboeck}},
  \bibinfo{journal}{Biological cybernetics} \textbf{\bibinfo{volume}{60}},
  \bibinfo{pages}{121} (\bibinfo{year}{1988}).

\bibitem[{\citenamefont{Gray et~al.}(1989)\citenamefont{Gray, K{\"o}nig, Engel,
  and Singer}}]{gray1989oscillatory}
\bibinfo{author}{\bibfnamefont{C.~M.} \bibnamefont{Gray}},
  \bibinfo{author}{\bibfnamefont{P.}~\bibnamefont{K{\"o}nig}},
  \bibinfo{author}{\bibfnamefont{A.~K.} \bibnamefont{Engel}}, \bibnamefont{and}
  \bibinfo{author}{\bibfnamefont{W.}~\bibnamefont{Singer}},
  \bibinfo{journal}{Nature} \textbf{\bibinfo{volume}{338}},
  \bibinfo{pages}{334} (\bibinfo{year}{1989}).

\bibitem[{\citenamefont{Wiesenfeld et~al.}(1998)\citenamefont{Wiesenfeld,
  Colet, and Strogatz}}]{PhysRevE.57.1563}
\bibinfo{author}{\bibfnamefont{K.}~\bibnamefont{Wiesenfeld}},
  \bibinfo{author}{\bibfnamefont{P.}~\bibnamefont{Colet}}, \bibnamefont{and}
  \bibinfo{author}{\bibfnamefont{S.~H.} \bibnamefont{Strogatz}},
  \bibinfo{journal}{Phys. Rev. E} \textbf{\bibinfo{volume}{57}},
  \bibinfo{pages}{1563} (\bibinfo{year}{1998}),
  \urlprefix\url{https://link.aps.org/doi/10.1103/PhysRevE.57.1563}.

\bibitem[{\citenamefont{Buck and Buck}(1968)}]{buck1968mechanism}
\bibinfo{author}{\bibfnamefont{J.}~\bibnamefont{Buck}} \bibnamefont{and}
  \bibinfo{author}{\bibfnamefont{E.}~\bibnamefont{Buck}},
  \bibinfo{journal}{Science} \textbf{\bibinfo{volume}{159}},
  \bibinfo{pages}{1319} (\bibinfo{year}{1968}).

\bibitem[{\citenamefont{Strogatz}(2003)}]{strogatz2003sync}
\bibinfo{author}{\bibfnamefont{S.~H.} \bibnamefont{Strogatz}},
  \emph{\bibinfo{title}{Sync: The Emerging Science of Spontaneous Order}}
  (\bibinfo{publisher}{Hyperion}, \bibinfo{year}{2003}).

\bibitem[{\citenamefont{Winfree}(1967)}]{winfree1967biological}
\bibinfo{author}{\bibfnamefont{A.~T.} \bibnamefont{Winfree}},
  \bibinfo{journal}{Journal of theoretical biology}
  \textbf{\bibinfo{volume}{16}}, \bibinfo{pages}{15} (\bibinfo{year}{1967}).

\bibitem[{\citenamefont{Winfree}(2001)}]{winfree2001geometry}
\bibinfo{author}{\bibfnamefont{A.~T.} \bibnamefont{Winfree}},
  \emph{\bibinfo{title}{The geometry of biological time}},
  vol.~\bibinfo{volume}{12} (\bibinfo{publisher}{Springer Science \& Business
  Media}, \bibinfo{year}{2001}).

\bibitem[{\citenamefont{Kuramoto}(1975)}]{kuramoto1975self}
\bibinfo{author}{\bibfnamefont{Y.}~\bibnamefont{Kuramoto}}, in
  \emph{\bibinfo{booktitle}{International symposium on mathematical problems in
  theoretical physics}} (\bibinfo{organization}{Springer},
  \bibinfo{year}{1975}), pp. \bibinfo{pages}{420--422}.

\bibitem[{\citenamefont{Acebr{\'o}n et~al.}(2005)\citenamefont{Acebr{\'o}n,
  Bonilla, Vicente, Ritort, and Spigler}}]{acebron2005kuramoto}
\bibinfo{author}{\bibfnamefont{J.~A.} \bibnamefont{Acebr{\'o}n}},
  \bibinfo{author}{\bibfnamefont{L.~L.} \bibnamefont{Bonilla}},
  \bibinfo{author}{\bibfnamefont{C.~J.~P.} \bibnamefont{Vicente}},
  \bibinfo{author}{\bibfnamefont{F.}~\bibnamefont{Ritort}}, \bibnamefont{and}
  \bibinfo{author}{\bibfnamefont{R.}~\bibnamefont{Spigler}},
  \bibinfo{journal}{Reviews of modern physics} \textbf{\bibinfo{volume}{77}},
  \bibinfo{pages}{137} (\bibinfo{year}{2005}).

\bibitem[{\citenamefont{Nozaki et~al.}(1999)\citenamefont{Nozaki, Mar, Grigg,
  and Collins}}]{PhysRevLett.82.2402}
\bibinfo{author}{\bibfnamefont{D.}~\bibnamefont{Nozaki}},
  \bibinfo{author}{\bibfnamefont{D.~J.} \bibnamefont{Mar}},
  \bibinfo{author}{\bibfnamefont{P.}~\bibnamefont{Grigg}}, \bibnamefont{and}
  \bibinfo{author}{\bibfnamefont{J.~J.} \bibnamefont{Collins}},
  \bibinfo{journal}{Phys. Rev. Lett.} \textbf{\bibinfo{volume}{82}},
  \bibinfo{pages}{2402} (\bibinfo{year}{1999}),
  \urlprefix\url{https://link.aps.org/doi/10.1103/PhysRevLett.82.2402}.

\bibitem[{\citenamefont{T{\"o}njes}(2010)}]{tonjes2010synchronization}
\bibinfo{author}{\bibfnamefont{R.}~\bibnamefont{T{\"o}njes}},
  \bibinfo{journal}{Physical Review E} \textbf{\bibinfo{volume}{81}},
  \bibinfo{pages}{055201} (\bibinfo{year}{2010}).

\bibitem[{\citenamefont{Bag et~al.}(2007)\citenamefont{Bag, Petrosyan, and
  Hu}}]{bag2007influence}
\bibinfo{author}{\bibfnamefont{B.~C.} \bibnamefont{Bag}},
  \bibinfo{author}{\bibfnamefont{K.}~\bibnamefont{Petrosyan}},
  \bibnamefont{and} \bibinfo{author}{\bibfnamefont{C.-K.} \bibnamefont{Hu}},
  \bibinfo{journal}{Physical Review E} \textbf{\bibinfo{volume}{76}},
  \bibinfo{pages}{056210} (\bibinfo{year}{2007}).

\bibitem[{\citenamefont{Moro and S{\'{a}}nchez}(1998)}]{Moro_1998}
\bibinfo{author}{\bibfnamefont{E.}~\bibnamefont{Moro}} \bibnamefont{and}
  \bibinfo{author}{\bibfnamefont{A.}~\bibnamefont{S{\'{a}}nchez}},
  \bibinfo{journal}{Europhysics Letters ({EPL})} \textbf{\bibinfo{volume}{44}},
  \bibinfo{pages}{409} (\bibinfo{year}{1998}).

\bibitem[{\citenamefont{H{\"a}nggi}(1995)}]{hanggi1995p}
\bibinfo{author}{\bibfnamefont{P.}~\bibnamefont{H{\"a}nggi}},
  \bibinfo{journal}{Adv. Chem. Phys.} \textbf{\bibinfo{volume}{89}},
  \bibinfo{pages}{239} (\bibinfo{year}{1995}).

\bibitem[{\citenamefont{Jung and H\"anggi}(1987)}]{praucna}
\bibinfo{author}{\bibfnamefont{P.}~\bibnamefont{Jung}} \bibnamefont{and}
  \bibinfo{author}{\bibfnamefont{P.}~\bibnamefont{H\"anggi}},
  \bibinfo{journal}{Phys. Rev. A} \textbf{\bibinfo{volume}{35}},
  \bibinfo{pages}{4464} (\bibinfo{year}{1987}),
  \urlprefix\url{https://link.aps.org/doi/10.1103/PhysRevA.35.4464}.

\bibitem[{\citenamefont{Maggi et~al.}(2015)\citenamefont{Maggi, Marconi, Gnan,
  and Di~Leonardo}}]{mucna}
\bibinfo{author}{\bibfnamefont{C.}~\bibnamefont{Maggi}},
  \bibinfo{author}{\bibfnamefont{U.~M.~B.} \bibnamefont{Marconi}},
  \bibinfo{author}{\bibfnamefont{N.}~\bibnamefont{Gnan}}, \bibnamefont{and}
  \bibinfo{author}{\bibfnamefont{R.}~\bibnamefont{Di~Leonardo}},
  \bibinfo{journal}{Scientific Reports} \textbf{\bibinfo{volume}{5}},
  \bibinfo{pages}{10742 EP } (\bibinfo{year}{2015}),
  \urlprefix\url{https://doi.org/10.1038/srep10742}.

\bibitem[{\citenamefont{Sakaguchi}(1988)}]{sakaguchi1988cooperative}
\bibinfo{author}{\bibfnamefont{H.}~\bibnamefont{Sakaguchi}},
  \bibinfo{journal}{Progress of theoretical physics}
  \textbf{\bibinfo{volume}{79}}, \bibinfo{pages}{39} (\bibinfo{year}{1988}).

\bibitem[{\citenamefont{Risken}(1996)}]{risken1996fokker}
\bibinfo{author}{\bibfnamefont{H.}~\bibnamefont{Risken}}, in
  \emph{\bibinfo{booktitle}{The Fokker-Planck Equation}}
  (\bibinfo{publisher}{Springer}, \bibinfo{year}{1996}), pp.
  \bibinfo{pages}{63--95}.

\bibitem[{\citenamefont{Marconi et~al.}(2017)\citenamefont{Marconi, Sarracino,
  Maggi, and Puglisi}}]{marconi2017self}
\bibinfo{author}{\bibfnamefont{U.~M.~B.} \bibnamefont{Marconi}},
  \bibinfo{author}{\bibfnamefont{A.}~\bibnamefont{Sarracino}},
  \bibinfo{author}{\bibfnamefont{C.}~\bibnamefont{Maggi}}, \bibnamefont{and}
  \bibinfo{author}{\bibfnamefont{A.}~\bibnamefont{Puglisi}},
  \bibinfo{journal}{Physical Review E} \textbf{\bibinfo{volume}{96}},
  \bibinfo{pages}{032601} (\bibinfo{year}{2017}).

\bibitem[{\citenamefont{Januszewski and
  Kostur}(2010)}]{januszewski2010accelerating}
\bibinfo{author}{\bibfnamefont{M.}~\bibnamefont{Januszewski}} \bibnamefont{and}
  \bibinfo{author}{\bibfnamefont{M.}~\bibnamefont{Kostur}},
  \bibinfo{journal}{Computer Physics Communications}
  \textbf{\bibinfo{volume}{181}}, \bibinfo{pages}{183} (\bibinfo{year}{2010}).

\bibitem[{\citenamefont{Paoluzzi et~al.}(2018)\citenamefont{Paoluzzi, Marconi,
  and Maggi}}]{PRExy}
\bibinfo{author}{\bibfnamefont{M.}~\bibnamefont{Paoluzzi}},
  \bibinfo{author}{\bibfnamefont{U.~M.~B.} \bibnamefont{Marconi}},
  \bibnamefont{and} \bibinfo{author}{\bibfnamefont{C.}~\bibnamefont{Maggi}},
  \bibinfo{journal}{Phys. Rev. E} \textbf{\bibinfo{volume}{97}},
  \bibinfo{pages}{022605} (\bibinfo{year}{2018}),
  \urlprefix\url{https://link.aps.org/doi/10.1103/PhysRevE.97.022605}.

\bibitem[{\citenamefont{Paoluzzi et~al.}(2016)\citenamefont{Paoluzzi, Maggi,
  Marini Bettolo~Marconi, and Gnan}}]{PREphi4}
\bibinfo{author}{\bibfnamefont{M.}~\bibnamefont{Paoluzzi}},
  \bibinfo{author}{\bibfnamefont{C.}~\bibnamefont{Maggi}},
  \bibinfo{author}{\bibfnamefont{U.}~\bibnamefont{Marini Bettolo~Marconi}},
  \bibnamefont{and} \bibinfo{author}{\bibfnamefont{N.}~\bibnamefont{Gnan}},
  \bibinfo{journal}{Phys. Rev. E} \textbf{\bibinfo{volume}{94}},
  \bibinfo{pages}{052602} (\bibinfo{year}{2016}),
  \urlprefix\url{https://link.aps.org/doi/10.1103/PhysRevE.94.052602}.

\bibitem[{\citenamefont{Farage et~al.}(2015)\citenamefont{Farage, Krinninger,
  and Brader}}]{farage2015effective}
\bibinfo{author}{\bibfnamefont{T.~F.} \bibnamefont{Farage}},
  \bibinfo{author}{\bibfnamefont{P.}~\bibnamefont{Krinninger}},
  \bibnamefont{and} \bibinfo{author}{\bibfnamefont{J.~M.}
  \bibnamefont{Brader}}, \bibinfo{journal}{Physical Review E}
  \textbf{\bibinfo{volume}{91}}, \bibinfo{pages}{042310}
  (\bibinfo{year}{2015}).

\bibitem[{\citenamefont{Marconi et~al.}(2016)\citenamefont{Marconi, Paoluzzi,
  and Maggi}}]{marconi2016effective}
\bibinfo{author}{\bibfnamefont{U.~M.~B.} \bibnamefont{Marconi}},
  \bibinfo{author}{\bibfnamefont{M.}~\bibnamefont{Paoluzzi}}, \bibnamefont{and}
  \bibinfo{author}{\bibfnamefont{C.}~\bibnamefont{Maggi}},
  \bibinfo{journal}{Molecular Physics} \textbf{\bibinfo{volume}{114}},
  \bibinfo{pages}{2400} (\bibinfo{year}{2016}).

\bibitem[{\citenamefont{Cates and Tailleur}(2015)}]{cates2015motility}
\bibinfo{author}{\bibfnamefont{M.~E.} \bibnamefont{Cates}} \bibnamefont{and}
  \bibinfo{author}{\bibfnamefont{J.}~\bibnamefont{Tailleur}},
  \bibinfo{journal}{Annu. Rev. Condens. Matter Phys.}
  \textbf{\bibinfo{volume}{6}}, \bibinfo{pages}{219} (\bibinfo{year}{2015}).

\end{thebibliography}

\end{document}